\documentstyle[sprocl]{article}
\begin{document}
\begin{flushright}
UCLA/99/TEP/13\\
April 1999
\end{flushright}

\title{Neutrinos, Cosmology and Astrophysics\footnote{To appear in the Proceedings of the International Workshop on Weak Interaction and Neutrinos (WIN99), Cape Town, South Africa, 24-30 Jan. 1999}}
\author{G. Gelmini}
\address{Department of Physics and Astronomy, UCLA, Los Angeles, 
CA 90095-1547}
\maketitle
\begin{abstract}
We review bounds on neutrino properties, in particular on their masses, coming mostly from cosmology, and also from astrophysics. 

\end{abstract}

Excluding laboratory experiments looking for neutrino oscillations and
observations of solar and atmospheric neutrinos (which have been covered
by other talks in this conference), most of our knowledge of neutrinos
comes from direct mass searches, cosmology and astrophysics.  We will deal
with these three areas in this order, concentrating our attention mostly on cosmological bounds. 
These come from data on the content, expansion
rate and lifetime of the Universe, the spectrum and anisotropies of the
Cosmic Microwave Background Radiation (CMBR), the large-scale structure (LSS)
of the Universe and the primordial abundance of the light elements (at present mostly from
deuterium) produced during nucleosynthesis (NS) in the early Universe.  There are lots of
quality data in all these areas.  We are entering into the ``precision era"
of cosmology, when most relevant cosmological parameters will become known
to a few percent.

\section{Direct Neutrino Mass Searches} 

From experiments we know that $m_{\nu_e} < {\rm few~eV}$.  Even if
``formal" upper limits of 5 eV (95\% C.L.) \cite{Mainz} and 1.7 eV
(95\% C.L.) \cite{Troitsk} have been given in 1998, they are obtained with
negative measured square masses, what points to a systematic error.  In fact
the Particle Data Book gives an upper bound of 15 eV \cite{PDB}.  The bounds
on the effective Majorana $\nu_e$ mass from $\beta\beta o\nu$ decays are of
about 1 eV.  The Heidelberg-Moscow $^{76}$Ge experiment quoted an upper
bound of 0.2 eV \cite{Heidelberg}, but there are still uncertanties related to 
the evaluations of the nuclear form factors involved.
There has been no recent change in the upper bound of the $\nu_\mu$ mass,
$m_{\nu_\mu} < 0.17$ MeV \cite{PDB}, while for $\nu_\tau$ a preliminary analysis of data from
LEP, combining results of ALEPH and OPAL gives $m_{\nu_e} < 15~{\rm MeV}$
(95\% C.L.) \cite{Cerutti}.  ALEPH and OPAL separately quoted upper
bounds of 18.2 MeV (95\% C.L.) and 27.6 MeV (95\% C.L.) respectively, while
new results of CLEO II give $m_{\nu_\tau} < 30~{\rm MeV}$ (95\% C.L.) \cite{CLEOII}. 

 Better bounds than these are derived from cosmological arguments if neutrinos are stable, and, if neutrinos are unstable, cosmology and astrophysics give bounds on 
their lifetimes and decay modes.

\section{Expansion, Age and Content of the Universe.}

The Hot Big Bang (BB), the standard model of cosmology, establishes that the
Universe is homogeneous, isotropic and expanding from a state of extremely
high temperature $T$ and density $\rho$.  The Hubble parameter $H$ (constant in space
but not in time) provides the proportionality between the velocity of recession
$v$ of faraway objects and their relative distance $d$, $v=Hd$ and $H=h$
100 km/sec Mpc.  Most observational determination are converging to
$h=0.65 \pm 0.15$ \cite{h} (for example, $h= 0.67\pm 0.12$ comes from type Ia
Supernovae \cite{h} while $0.72 \pm 0.17$ comes from combining cepheids
studies and other methods \cite{Freedman98}), for the present value of the
expansion parameter.  The lifetime of the Universe is counted from the
moment the expansion started, taken to be $t=0$.  The cooling of white dwarfs
provides a lower bound to the present age of the Universe of $t_o > 10$ Gyr \cite{dwarfs} and the age of the oldest globular clusters gives $t_o = 13-25$ Gyr ($t_o = 11.5 \pm 1.3~{\rm Gyr}$  \cite{Chaboyer}, plus 1-2 Gyr for the formation of the
galaxy where the globular cluster is.  The expansion rate, lifetime and the
content of the Universe are not independent. 
 In fact, $Ht_o = (h/0.75)(t_o/13$
Gyr) is a function of the densities of matter, radiation and vacuum in the
Universe.  For an empty Universe $Ht_o=1$.  The gravitational attraction of
matter and radiation slows down the expansion, so that $Ht_o<1$ in a matter
or radiation dominated Universe. In a vacuum dominated Universe $Ht_o>1$ instead, because gravitation is repulsive.  In synthesis, a larger $H$ implies
a shorter $t_o$ and decreasing the matter or radiation content or increasing
the vacuum content of the Universe makes $t_o$ longer.

Densities $\rho_i$ are usualy given in units of the density of a spatially
flat Universe, the critical density $\rho_c = 10.5~{\rm h}^2~{\rm (keV/cm)}^3$
= 1.88 10$^{-29}~ h^2$ (g/cm$^3$) as $\Omega_i = \rho_i/\rho_c$.  We call
$\rho_r$, the density of radiation (photons and other relativistic
particles) and $\rho_m$ the density matter (non-relativistic particles).
We will call here $\Omega_o = \Omega_m + \Omega_r$.
The vacuum energy provides a cosmological constant $\Lambda$, such that
$\rho_{\rm vac} = \Lambda/8\pi{\rm G} = \rho_\Lambda$. In our notation the total density of the Universe is $\Omega = \Omega_o + \Omega_\Lambda$.
  Matter is much more
abundant than radiation at present, thus $\Omega_o \simeq
\Omega_m$.  For a flat 
matter dominated Universe,
$Ht_o = 2/3$.  This means (using the relation given above) that 
$t_o\geq 13$ Gyr (necesary to accomodate globular clusters) requires
$h \leq 0.50$, a very low value of $h$.  Even using the absolute lower bound
$t_o \geq 10$ Gyr, we obtain $h \leq 0.65$, still values of h lower than many
present determinations.  If $h$ is actually larger than 0.65, then we live in
a Universe with a non-zero cosmological constant or open or both.  This tension
between $h$ and $t_o$ was until recently called the ``age crisis", but this
is not called a ``crisis" any longer.

 One speaks of a ``crisis" only when a
paradigm is challenged.  This paradigm was that of a flat matter dominated
Universe with $\Lambda=0$, until recently the model preferred by most
cosmologists due to its simplicity and aesthetic appeal.  This paradigm has
now been changed, mostly by the Type Ia Supernovae (SN) data, which point to a
non-zero value of $\Lambda$, and, to a lesser extent, by data on the LSS of the Universe, which suggests $\Omega_m < 1$.  Type Ia SN are white dwarfs which accrete mass from a comparison star and explode when reaching the Chandrasekar limit of about  1.4 solar masses, i.e. the maximum mass that can be supported
by the pressure of degenerate electrons.  Two different groups \cite{Perlmutter9798} \cite{Garnavich} using Type Ia SN as ``calibrated" candles
(i.e. objects of known intrinsic luminosity) measured the curvature of the
relation between distances and velocities (of which the linear term is given
by the present value of the Hubble parameter $H$).  Recession velocities are actually translated
into redshifts $z$.  Thus, distances $d$ are given by $d=H^{-1}[z+O(z^2)]$.  
At $z\simeq 0.5$, where most of the SN used are, the coefficient of the $z^2$ term depends
on the linear combination $\Omega_\Lambda - \Omega_m$ \cite{white}, and, in fact, the
confidence region given by both groups \cite{Perlmutter9798} \cite{Garnavich} lies along the line $\Omega_\Lambda-\Omega_m \simeq 0.5$,
in the $(\Omega_m,\Omega_\Lambda)$ plane.  With 42 high-redshift supernovae,
Perlmutter {\it et al.} \cite{Perlmutter9798} found a non-zero positive cosmological constant with
probability larger than 99\% (0.9979 in their primary fit).  In order to
determine $\Omega_\Lambda$ and $\Omega_m$ separately, one would wish to
combine this result with that of a complementary technique sensitive to a
different linear combination of the two quantities. In fact, an almost orthogonal linear combination is provided by the position of the 
first acoustic peak
in the multipole expansion of the CMBR anisotropy, which depends on the total
energy density of the Universe, $\Omega= \Omega_\Lambda + \Omega_m$ (see below).  Data already show
that this sum is not very different from one.  
The crossing of $\Omega_\Lambda -
\Omega_m \simeq 0.5$ and $\Omega_\Lambda + \Omega_m \simeq 1$, suggests the values $\Omega_m \simeq 0.3$ and $\Omega_\Lambda \simeq 0.7$ 
  (which would
saturate an earlier upper bound $\Omega_\Lambda < 0.7$ obtained for a flat
Universe from the frequency of gravitational lensing 
of quasars \cite{quasars}).
In a few years,
with better data,   $\Omega_m$ and $\Omega_\Lambda$ will be determined in this way to within 10\%. The satellite MAP, will be
launched by NASA in the year 2000 to measure the anisotropy of the CMBR (a European satellite, the Planck Surveyor, is expected for 2007).

The CMBR provides a snapshot of the Universe at the moment of recombination,
$t_{rec} \simeq 3\times 10^5$y, when atoms first became a stable.  Photons, which
had a very short mean free path in the ionized matter before recombination,
interact for the last time and reach us from that ``surface of last scattering".
This radiation has the best black-body spectrum in the Universe with deviation
of less than 0.005\% and a temperature $T = 2.7277^\circ$K measured by the
COBE (Cosmic Background Explorer) Satellite \cite{Fixsen}.  This radiation is
remarkably isotropic.  Anisotropies are due to our motion with
respect to the CMBR rest frame (which generates a dipole anisotropy) and due to
the density inhomogeneities that triggered structure formation in the Universe.
Results from COBE and other experiments in balloons show temperature
anisotropies $(\delta T/T) \equiv (T-\bar T)/\bar T < 10^{-4}$, where $\bar T$
is the average temperature (given above).  Once within the horizon, the
primordial density perturbations (in dark matter) set up sound or acoustic
oscillations in the fluid formed by photons, electrons and baryons before
recombination.   In the surface of last scattering the peaks of 
compression and rarefaction (for scales that are caught at extrema of their
oscillations) are seen as hot and cold spots respectively,  both of which 
appear as peaks in
the multipole expansion of the power spectrum of CMBR anisotropies.  The
horizon size at recombination corresponds to an angle $\theta_H \simeq  1^\circ  \sqrt{\Omega}$ in the present sky.  This apparent angular size 
depends on the geometry of the Universe: it is larger for a closed Universe
$(\Omega >1)$ and smaller in an open one $(\Omega < 1)$ \cite{Jungman}.
In multipole number $\ell \simeq 200^\circ/\theta$, the position of the 
horizon is at  $\ell_H \simeq (200 / \sqrt{\Omega})$.  Angles larger 
than $\theta_H$,  $\ell < \ell_H$, correspond
to causally disconnected regions at the time of emission of the CMBR 
photons, where no acoustic oscillations could have been set.  Only values
$\ell \geq \ell_H$ correspond to scales within the horizon.  Thus the
position of the first acoustic peak (a compression peak) should happen at
$\ell_H$, which depends on $\Omega$.  Present data on
anisotropies, even if not conclusive, show the first peak at $\ell$'s not much
lower than 100 or higher than 300.

\section{Cosmic Energy Densities}

Maybe the most important cosmological constraint on stable neutrinos is the
mass bound coming from their cosmic energy density.  In the same way we have
a cosmic background of photons, we expect the existence of a yet never seen
cosmic background of relic neutrinos.  Knowing so well the CMBR temperature,
we know with great accuracy the number and energy density of the CMBR photons
which are the most abundant in the Universe by several orders of magnitude,
$n_\gamma = 2\zeta(3) T^3/\pi^2 = 412/{\rm cm}^3$, $\rho_\gamma =
\pi^2T^4/15 = 4.71\times 10^{-34}~({\rm g/cm}^3)$.  We can compute the
expected abundance of neutrinos relative to photons.  For light standard neutrinos (of mass $m_\nu < 1$ MeV and  no lepton number asymmetry)
$n_{\nu_i} + n_{\bar\nu_i} = (3/11) n_\gamma = 112/{\rm cm}^3$ (including both
neutrinos and antineutrinos in equal numbers) for each light neutrino species.
The temperature of neutrinos is lower than that of photons 
$T_\nu = (4/11)^{1/3} T = 1.9^\circ{\rm K} = 1.6~10^{-4}~{\rm eV}$.  Knowing
$T_\nu$, we can compute the contribution of each relativistic  $\nu$-species  to the present radiation energy, usually parametrized as
$\rho_{\rm rad} = (\pi^2/30)~g_*(T)~ T^4$, where $g_*(T)$ is the effective
number of relativistic degrees of freedom.  Every standard neutrino species
(with no lepton number asymmetry) adds 0.454 to $g^*$, while photons 
contribute with 2.  Photons and three relativistic neutino species add up to
$g_* = 3.362$ and $\Omega_{\rm rad} h^2 \simeq 4\times 10^{-5}(g_*/3.36)$.
If one or more standard neutrino species are non-relativistic at present
$(m_{\nu_i}>T_\nu)$, then their contribution to the present density of the
Universe is $\rho_\nu = \sum_i m_{\nu_i}(n_{\nu_i} + n_{\bar\nu_i}) =
\Omega_\nu\rho_c$, thus
\begin{equation}
\Omega_\nu h^2 = \frac{\sum_i m_{\nu_i}}{92~{\rm eV}}~.
\end{equation}
Only left-handed (non-relativistic) neutrinos (with no lepton number asymmetry) are considered here (for
Dirac neutrino masses $< 1~{\rm keV}$ this is correct, because the
contribution of the right-handed states is negligible).  If $m_{\nu_i} =
0.4~h~{\rm eV}$ (0.3 eV with $h$ = 0.65) standard neutrinos would be as abundant as
luminous matter, namely the matter associated with typical stellar populations,
which is $\Omega_{\rm lum} \simeq 0.004~h^{-1}$ (0.6 $10^{-2}$ for $h$ = 0.65).
This matter is baryonic, but Big Bang Nucleosynthesis (BBNS) arguments 
emply that the total density of baryons, $\Omega_B$, is larger.  Estimates based on the sole
density of $D$ \cite{Burles}, whose primordial abundance is the best
known among the light elements \cite{Tytler}, give $\Omega_B = 
(0.019 \pm 0.0024)~h^{-2}$, comparable to the density of a standard 2 eV
neutrino.    Due to uncertainties in the observational
upper bound on the abundance $^4$He, only $D$ is used to obtain this range.
Also due to this uncertainty the limit on $N_\nu$,
the number of equivalent standard neutrino
species in equilibrium during NS, is uncler.  At present
there are two estimates of primordial $^4$He.  The lowest one,\cite{Olive} together
with data on $D$ would require $N_\nu < 3$, while the higher \cite{Izotov},
with a prior $N_\nu > 3$, implies $N_\nu < 3.2$ (95\% C.L.) \cite{Burles}.

The gravitationally dominant mass component of the Universe is ``dark", i.e.
it is not seen either in emission or absorption of any type of electromagnetic
radiation.  This is the dark matter (DM).  Recent measurements give
$\Omega_{\rm DM} > 0.15$ (an absolute lower bound, coming from satellites of
spiral galaxies), $\Omega_{\rm DM} = (0.19 \pm 0.06)$ B (with B $\simeq$ 1,
from the mass over light ratio of clusters), $\Omega_{\rm DM} = 0.44 \pm 0.11$
(from the baryon fraction in clusters, using BBNS), $\Omega_{\rm DM} \simeq
0.55 \pm 0.17$ (from the abundance of high-$z$ clusters).  Notice
that none of these dynamical estimates reaches 1.  Important for neutrinos
is a bound that depends on the total density of DM, coming from structure
formation arguments (presented in the following section), that say that most of the DM in the Universe
should be cold (i.e. non relativistic at temperatures of about $T \simeq 1$ keV, when galaxies should start forming), called CDM, and
only a small amount could be hot 
(i.e. relativistic at $T \simeq 1$ keV, such as light neutrinos), called HDM. This bound is
$\Omega_\nu \leq (0.2-0.3)\Omega_{\rm m}$, where
 $\Omega_{\rm m} \simeq \Omega_{\rm DM} = \Omega_{\rm CDM}+ \Omega_\nu$.  With $\Omega_{\rm m} = 1$,
$\Omega_\nu \leq 0.2$ gives an often quoted bound $\sum_i m_{\nu_i} \leq
5~{\rm eV}$, for h = 0.5 (this low value of $h$ is necessary in a flat matter
dominated Universe to account for the age of the Universe, as mentioned
above).  However, lower values of $\Omega_{\rm DM}$, as measurements seen now to point to, would lead to more 
stringent bounds on $\Omega_\nu$. 

We have spoken so far about neutrinos with no or negligible leton number asymmetry. However lepton asymmetries in neutrinos may be large (see the end of Sec. 4).

\section{Structure Formation of the Universe.}
The Universe looks lumpy at scales $\lambda \simeq 100$ Mpc, we see galaxies,
clusters, superclusters, voids, walls.  But it was very smooth at the surface
of last scattering of the CMBR  and later.  
Inhomogeneities have been seen as anisotropies in the CMBR, so
 the density contrast $\delta\rho/\rho \equiv (\rho(x)-\rho)/\rho$ (where $\rho$ is the average density) 
 cannot be much larger than $\delta T/T \simeq 10^{-4}$.
So inhomogeneities in density start small and grow through the Jeans (or
gravitational) instability; gravitation tends to further empty underdense
regions and to further increase the density of overdense regions.  One can
follow analytically the evolution due to gravity of the density contrast
in the linear regime, where $\delta\rho/\rho < 1$.  In a static fluid the rate
of growth of $\delta\rho/\rho$ is exponential, but in the Universe (an
expanding fluid) it slows down into either a power law, $\delta\rho/\rho 
\sim a(t)$, in a matter dominated Universe, or it stops, 
$\delta\rho/\rho \simeq$ constant, in a radiation or a curvature dominated
Universe (a matter dominated open Universe becomes curvature dominated for
$a(t) \geq \Omega_o/(1-\Omega_o)$). Here $a(t)$ is the scale factor of the Universe, which accounts for the Hubble expansion of the linear
dimensions of the Universe.
Perturbations have different  physical linear dimensions $\lambda = a(t) \lambda_{\rm com}$, where $\lambda_{\rm com}$ are  
linear dimensions measured in comoving coordinates (those 
that expand with the Hubble flow). With the usual choice of $a = 1$ 
at present, $\lambda_{\rm {com}}$ are the present actual linear dimensions.  
Since $a\sim t^\alpha$ with $\alpha < 1$  while 
the horizon $ct$ grows linearly with $t$, the horizon increases 
with time even in comoving coordinates, encompassing more material as
time goes. When $\lambda = ct$ we say
the perturbation of size $\lambda$ ``enters" into the horizon, we could better 
say the perturbation is first encompassed by the horizon.  This moment is
called ``horizon-crossing" and it happens at different times for different
linear scales $\lambda$, larger scales cross later.

Independently of the origin of the primordial fluctuation, it is convenient
to specify the spectrum of fluctuations at horizon-crossing,
$(\delta\rho/\rho)_{\rm hor}$.  A spectrum scale
invariant at horizon-crossing, namely with $(\delta\rho/\rho)_{\rm hor}$ =
constant, is called a Harrison-Zel'dovich spectrum.  COBE observations have shown the spectrum at horizon-crossing is in fact scale
invariant or very close to it.

After horizon-crossing, physical interactions act upon the inhomogeneities
and generate a ``processed" spectrum, which determines which structures
are formed first.
This question  leads to the distinction
of three types of DM: hot (HDM), warm and cold (CDM) (i.e. relativistic, becoming non-relativistic  and non-relativistic at temperatures of order keV).

Simulations have shown that CDM must be the most abundant form of matter,
because the ``processed" spectrum of perturbations generated in standard
CDM models reproduces the observations within 10\%.  Standard CDM models,
make the simplest assumptions, namely $\Omega_{\rm CDM} + \Omega_{\rm B} \simeq
\Omega_o = 1,~\Lambda = 0$, scale invariant perturbations at horizon crossing,
and a scale independent ``biasing" by which only the highest peaks in the CDM
density distribution end up forming galaxies.  There is only one feature in
the processed spectrum of CDM perturbations, a change of slope at the present
scale that corresponds to the horizon at the moment of matter-radiation
equality, $\lambda_{\rm eq}$. 
The Universe is matter dominated at present, but due  to the different
evolution with temperature $T$ of the density of matter and radiation,
$\rho_{\rm m} \sim T^3$, $\rho_{\rm r} \sim T^4$, the radiation was
dominant in the past, at $T>T_{\rm eq}$, where $T_{\rm eq}$
is the temperature of matter-radiation equality $\rho_{\rm r}(T_{\rm eq}) =
\rho_{\rm m}(T_{\rm eq})$, $T_{\rm eq} \simeq 5.8 {\rm eV}
\Omega_oh^2(3.36/g_*)$.  $\Omega_o$ is the present matter density 
(neglecting the present small radiation contribution) and
$g_*$ is the number of effective relativistic degrees of freedom 
($g_* = 3.36$ with photons and three relativistic neutrino species).  The
present physical size of the horizon then is
\begin{equation}
\lambda_{\rm eq} \simeq 10~ {\rm Mpc}
\left(\frac{g_*}{3.36}\right)^{1/2} \frac{1}{\Omega_oh^2}~.
\end{equation}

Perturbations with $\lambda <
\lambda_{\rm eq}$ enter into the horizon at $t < t_{\rm eq}$, 
when the Universe is radiation dominated. 
 They cannot grow while the Universe is
radiation dominated, so they all start growing together at $t = t_{\rm eq}$
and they roughly have the same amplitude today, if they all start with the
same amplitude at horizon crossing. Perturbations with
$\lambda > \lambda_{\rm eq}$, instead,
enter into the horizon at $t > t_{\rm eq}$, when the Universe is matter
dominated, and, thus, start growing immediately.  Consequently, perturbations
at larger scales enter later, have less time to grow and their amplitude is
smaller at present.  Once $\lambda_{\rm eq}$ (the location of the change
slope) is fixed, the only remaining free parameter in the processed spectrum of CDM is an overall normalization,  provided by the CMBR anisotropy measured by COBE at
large scales, $\theta > 20^\circ$.  Density perturbations at these large
scales entered into the horizon very recently (so they did not grow much),
thus providing a measurement of $(\delta\rho/\rho)$ at horizon
crossing, $(\delta\rho/\rho)_{\rm hor}$
(for more details, see e.g. Ref. 21). 

 While both the shape and normalization so obtained are
almost right, they do not fit the observations \cite{Ostriker}. 
 The spectrum  of standard CDM models has too much
power on small scales (large k$\sim \lambda^{-1}$), the scales of galaxy clusters and smaller.

  Once the normalization given by COBE is fixed, there are
several possibilities to change the spectrum to agree with observations.
Because HDM tends to erase  structure at small scales (while neutrinos are relativistic)
one of the solutions consists in adding to the CDM a bit of HDM, namely
neutrinos,  in what are called  mixed DM (MDM) or hot-cold DM (HCDM) 
models \cite{Shafi}.
In particular, models with $\Omega_\nu=0.2$, what corresponds to
 $\sum_i {m_{{\nu}_{i}}}$ = 5 eV, and the rest of $\Omega_o$ in CDM plus
some baryons, with $\Omega = 1$, $\Lambda = 0$  and a scale invariant
spectrum of fluctuations at horizon crossing work well. However
other possible variations of the standard CDM models  also work well
to fit the LSS data, 
for example  that of a ``tilted''
primordial spectrum of fluctuations at horizon crossing, one that slightly 
favors larger scales over smaller scales
(instead of the flat, scale invariant, Harrison-Zel'dovich spectrum) within
the COBE observational limits.  This is called ``tilted'' CDM (TCDM).\cite{Cen}
 A mixed model with
 both some neutrinos and some ``tilt"
also does work, and in these models 
$\Omega_\nu < 0.2$ (see, for example,\cite{DGT}).

Another family of solutions is obtained by realizing that
a shift towards larger scales of the only feature in the CDM spectrum, i.e.
$\lambda_{\rm eq}$, given in Eq. (2), the scale where the
slope in the spectrum changes,
 is enough to provide good agreement with observations, since
it effectively amounts to increasing the power of the spectrum at scales 
larger than the break point
$(\lambda > \lambda_{\rm eq})$ with respect to those smaller  than it
$(\lambda < \lambda_{\rm eq})$.
Using Eq. (2) the relation
 $\lambda_{\rm eq} \equiv (10h^{-1}M_{\rm pc})\Gamma^{-1}$ defines
the ``shape parameter" \cite{Efstathiou}
$\Gamma \equiv \Omega_oh(g_*/3.36)^{-1/2}$.  The LSS
data require $\Gamma \simeq 0.25\pm 0.05$, while standard CDM models (with
the standard choices of $h = 0.5, \Omega_o = 1,~g_* = 3.36$) has $\Gamma =
0.05$.  In fact, as we have explained, a larger $\lambda_{\rm eq}$, thus a
smaller $\Gamma$, would provide agreement with data.  In order to lower the
value of $\Gamma$ with respect to that of the standard CDM model one needs to
either, 1): lower $h$, $h < 0.5$ \cite{Bartlett}
(what implies an older Universe and is very unlikely given the present determinations of $h$), or 2): increase
$g_*$ (namely increase the radiation content of the Universe at
$t_{\rm eq}$), or 3): lower $\Omega_o$ (i.e. take $\Omega_o < 1$), so that
we either live in an open Universe (open CDM models, OCDM) if
$\Lambda = 0$ or in a Universe with a cosmological constant that provides
$\Omega_{\rm vac} = 1-\Omega_o$ ($\Lambda$CDM models \cite{Turner}),  or 4: a combination of all three above.

A way of obtaining the large amount of radiation needed for the 
second possibility is through a heavy
neutrino decaying into relativistic particles, i.e. radiation,
 with the right combination of mass and
lifetime, in so-called $\tau$CDM models \cite{Bardeen} \cite{Dodel}.
 A massive neutrino matter
dominates the energy density of the Universe as soon as it becomes
non-relativistic, i.e. as soon as 
 $m_\nu \geq T$ (since $n_\nu \simeq n_\gamma$ and
$\rho_\nu = n_\nu m_\nu,~\rho_{\rm rad} \simeq n_\gamma T$), thus their decay
products radiation-dominate the Universe at decay.  For $m_\nu < 1$ MeV standard  neutrinos the 
right mass-lifetime combination lie on a narrow strip around the previously
mentioned ``galaxy formation" bound \cite{Hut}.
 Near this bound, at the boundary
between being irrelevant and harmful, unstable neutrinos could help in the
formation of structure in the Universe \cite{Bardeen}.
 A heavier  neutrino, of $m_\nu \simeq 1 - 10$ MeV, necessarily $\nu_\tau$,
decaying at or just before nucleosynthesis, $\tau = 0.1 - 100$ sec, 
would also provide a solution \cite{Dodel}.
  The $\nu_\tau$ decay modes involved here should all be into neutral
particles, $\nu_\tau \to 3~\nu'{\rm s}$ or $\nu_\tau \to \nu\phi$, with
$\phi$ a Majoron (a zero mass Goldstone boson) for example.  All visible
modes, i.e. producing electrons or photons, are forbidden in the necessary
range.

 Radiation-matter equality may also be delayed   
by the existence of large lepton asymmetries in neutrinos, so that these may be more abundant than photons \cite{Sarkar},
$(n_\nu/n_\gamma)> 1$ (and dominate the entropy $s$ of the Universe,
$n_\nu /s = O(1)$). Relic neutrinos this abundant would be Fermi-degenerate, since their chemical potential $\mu_\nu$
would be larger than their temperature.  Let us recall that, while charge neutrality imposes a lepton number
asymmetry in electrons as large as the baryon asymmetry in protons, i.e.
$(n_e-n_{\bar e})/n_\gamma\simeq 10^{-10}$, no such restrictive bound operates
on neutrinos. 
We will return below to this  very recently explored possiblity,
that we call LCDM (for CDM with large lepton asymmetry L).

All these modified CDM models seem to be able to fit present large scale structure data, however
they predict very different patterns of acoustic peaks in the CMBR 
anisotropy power spectrum, and already present data on this anisotropy allow to constrain the models. All recent analizes perform a combined fit to LSS and CMBR anisotropy data.
  CHDM \cite{Gawiser}, $\tau$CDM \cite{Jenkins}, and LCDM \cite{Sarkar} have been favorably compared with others in their ability
to fit LSS and CMBR data.  

 Gawiser and Silk \cite{Gawiser} claimed that CHDM with
5 eV total neutrino mass ($\Omega_\nu = 0.2$, $\Omega_m = 1$, $\Omega_\Lambda = 0$, $h = 0.5$) gives the best fit among all the 10 models they studied.
However, they mention that this model, as all others with $\Omega=\Omega_0=1$ (flat matter dominated Universe) may not account for the recent evidence for early galaxy formation.  In fact, evidence has been found for the
existence of a large amount of bright galaxies rather early, at redshifts
$z\simeq 3$, and Somerville, Primack and Faber \cite{Somerville} concluded that
no $\Omega =\Omega_o= 1$ model with a realistic power spectrum
makes anywhere near enough of them.  This can be understood by recalling that
in a matter-dominated universe $(\delta\rho/\rho$) grows as the scale
factor while in a curvature or $\Lambda$-dominated universe the growth of
$(\delta\rho/\rho)$ stops.  Thus, in order to get to the same present level
of structure, the density contrast of perturbations $(\delta\rho/\rho)$ should be bigger at early times in the
later case (growth of $(\delta\rho/\rho)$ stops at some point in the past)
than in the former (in which the growth of $(\delta\rho/\rho)$ continues 
until now).  Based on these considerations, as well as on the Type Ia SN data,
which favor $\Lambda > 0$, Primack and Gross \cite{Primack} studied the 
combined fit to LSS and CMBR data of a $\Lambda$CHDM model,namely flat models
with $\Omega_m < 1$ and  $\Omega_\Lambda= 1- \Omega_m$ , and with some HDM in neutrinos.  They found
the best fits had $\Omega_m$ = 0.5 (0.6) with ($\Omega_\nu/\Omega_m)$ = 
0.1 (0.2) (namely $\Omega_\nu$ = 0.05 (0.12)) 
corresponding to $\sum_i m_{\nu_i} = 1.6 (4)$ eV (since they used h = 0.6).  They found that the addition of HDM
does not change a $\Lambda$CDM model by much, contrary to the substantial
improvement this addition provided to standard CDM models.  However, they also
concluded that $\Lambda$CDM and also $\Lambda$CHDM models provide a
relatively poor fit to the LSS data (to the power spectrum near the peak),
what is also mentioned in Ref. 33.  The new large scale surveys
under way (2dF and SDSS) will be crucial in determining the viability
of these models. 

 Light neutrinos not being a necessary addition to the matter
composition of the Universe to explain the known data,  the studies
just mentioned provide an upper bound (already mentioned in Section 3)
on the relative amount of neutrinos with respect to CDM: in all of them
$\Omega_\nu/\Omega_m \leq 0.2-0.3$.  

Let us finally return to the possibilty of relic neutrinos with a large lepton 
asymmetry.   Adams and Sarkar \cite{Sarkar} found that a relic neutrino species with
chemical potential $\mu_\nu = 3.4~T_\nu$ added to a standard CDM model
(flat matter dominated universe) provides a good fit to the LSS and CMBR
data. The best bounds on large $\mu_\nu$ come from BBNS (which
becomes severely non-standard in the presence of large neutrino asymmetries)
and structure formation.  Relic neutrinos with large
chemical potentials $\mu_\nu > T$, would form a Fermi degenerate background
with number density
$n_\nu = (12 \zeta(3))^{-1} \left (T_\nu / T_\gamma \right )^3
[\pi^2[(\mu_\nu / T)+(\mu_\nu / T)^3] = 0.0252 [9.87 (\mu_\nu / T) + (\mu_\nu / T)^3]$,
where $(T_\nu/T_\gamma)^3$ has the standard value of (4/11) for
$(\mu_\nu/T) < 12$.  For $\mu/T = 3.4$, the density of neutrinos would
be $n_\nu = 1.8~n_\gamma = 756/{\rm cm}^3$.  Parenthetically let us point
out that slightly larger chemical potentials,
 still alowed by LLS and CMBR data, could make neutrinos with the
mass implied by SuperKamiokande data, in the case of hierarchial masses,
$m_\nu = \sqrt{\delta m^2} \simeq 0.07~{\rm eV}$, a relevant HDM component.
For example with $(\mu_\nu/T) = 4.6$, the relic neutrino density would be
$n_\nu = 3.6~n_\gamma$ and a neutrino of mass 0.7 eV would have
$\Omega_\nu h^2 = 0.01$.\cite{Gelmini}  These neutrinos would still be 
relativistic at the moment of radiation-matter equality, so they would have
on structure formation almost the same effect as the massless neutrinos
studied by Adams and Sarkar. The large  neutrino asymmetries necessary in 
these models could possibly be generated through neutrino oscillations after the elctroweak phase transition (this needs to be studied) and certainly with the Affleck-Dyne mechanism. \cite{Casas}

Many possiblities are still open, but the quality data necessary to
discriminate among models are coming, and a confirmation of one  of
them may be possible within a few to ten years.  Besides getting to know
$\Omega$, $\Omega_B$, $\Omega_\Lambda$, $\Omega_m$, H, $t_o$ etc., the relevant
cosmological parameters to a few \%, standard neutrinos with no lepton 
asymmetry will be seen either as HDM for $m_\nu \geq 1$ eV (may be up to
0.3 with low $\Omega_m$)\cite{Hu1} or for lighter neutrinos, the number of
neutrino species will be determined with precision similar to the NS
bounds.\cite{Hu2}  The case of large neutrino lepton asymmetry has yet to
be studied.

\section{Astrophysics.}

Neutrinos may have an important role in the evolution of some types of stars,
in particular in the explosion of Type II SN.  These are stars in which the
Fe-core reaches the Chandrasekar limit  and, thus,
collapses into a neutron star, trapping the emitted neutrinos for several
seconds within a region called ``neutrino-sphere", and exploding the mantle of the star. Actually the neutrino spheres of different neutrinos types are slightly different leading to different neutrino average energies.  Most of the binding energy of the remaining neutron star
goes into neutrinos. Thus,  the explosion mechanism is sensitive to non-standard neutrino properties.  Some examples are the following. 

A problem in Type II SN modelling is that
 the shock wave which should explode the mantle  stalles before getting to do it. This problem might be solved
if the $\nu_e$ arriving at the shock from the $\nu$-sphere are actually
$\nu_\mu$ or $\nu_\tau$ which oscillated resonantly into $\nu_e$
on their way.\cite{Fuller}   These $\nu_e$ arriving at the shock would have the higher energy ofthe originally emitted $\nu_\mu$-$\nu_\tau$, 24-27 MeV, instead of the lower energy of 10-12 MeV with which $\nu_e$'s are emitted, leading to a more efficient energy transfer to the shock. This would require $\Delta m^2 = 10^2$ to $10^4$eV$^2$ and mixing angles $\sin^2(2 \theta) > 10^{-7}$ approximately.  However, these resonant oscillations may prevent the
r-process (rapid neutron capture) synthesis of heavy elements, whose
preferred site is behind the shock wave of an exploding Type II SN. Because the 
energy of $\nu_e$ in this case would be larger than that of 
$\bar\nu_e$, 14-17 MeV, the environment would become proton rich, while r-process requires a
neutron rich medium.\cite{Qian} This would happen for  
$\Delta m^2 = 10$ to $10^4$eV$^2$ and mixing angles $\sin^2(2 \theta) > 10^{-6}$ approximately. The matter enhanced oscillations  of
$\nu_e$ and a sterile neutrino may instead help r-process, by eliminating
the $\nu_e$ which decimate neutrons (through the reaction $n\nu_e\to pe^-$) \cite{Fetter}, for $\Delta m^2 = 1$ to $10^2$eV$^2$ and mixing angles $\sin^2(2 \theta) > 10^{-3}$ approximately.  These arguments involving r-process in Type II SN are to
be aken with a grain of salt, considering that there is another possible site for it, in
binary neutron star - neutron star mergers. 

As the
last point, let us mention ``pulsar kicks".  Pulsars, which are neutron stars
with large magnetic fields, are not confined to the disk of our galaxy, where
they must have been born, but move in all directions with large velocities
of 500 km/sec on average.  A ``pulsar kick" which may impart these velocities may result from the existence of a
matter-enhanced resonant region between the more internal $\nu_\mu, \nu_\tau$ and the more external
$\nu_e$ neutrino spheres.\cite{Kusenko}  Along the direction of the magnetic
field the resonance happens at larger densities, thus more internally in the 
protostar.\cite{D'Olivo}  The $\nu_e$ oscillated into $\nu_\mu$ or $\nu_\tau$ within the $\nu_e$ sphere (from which they could  not escape otherwise) but outside the $\nu_\mu,\nu_\tau$ sphere can freely
escape, and they carry more energy away when the conversion happens the
farthest from the surface of the star (where temperatures are higher).  Thus,
the deformation of the resonance region due to the presence of a large
magnetic  field, has as a consequence the emission of hotter converted $\nu_e$
 neutrinos in the direction of
the magnetic field and cooler ones in the opposite direction.  A 1\% asymmetry
in the total momentum carried by the emitted neutrinos would be enough to
impart to the remaining neutron star the large velocities of pulsars.\cite{Kusenko}  Several versions of this mechanism have been proposed.\cite{others}

\noindent{\it Acknowledgments}

I thank the organizers of this workshop for their invitation.
This work was supported in part by the U.S. Department of Energy under Grant
DE-FG03-91ER 40662 Task C.

\end{document}